\documentclass[11pt]{article}
\usepackage{amsmath,amsfonts,amssymb}
\numberwithin{equation}{section}

\newcommand{\nc}{\newcommand}
\nc{\bib}{\bibitem}
\nc{\al}{\alpha}
\nc{\g}{\gamma}
\nc{\G}{\Gamma}
\nc{\D}{\Delta}
\nc{\eps}{\epsilon}
\nc{\la}{\lambda}
\nc{\La}{\Lambda}
\nc{\var}{\varphi}
\nc{\pa}{\partial}
\nc{\nn}{\nonumber \\ }
\nc{\be}{\begin{equation}}
\nc{\ee}{\end{equation}}
\nc{\bea}{\begin{eqnarray}}
\nc{\eea}{\end{eqnarray}}
\nc{\bra}[1]{\langle {#1}|}
\nc{\ket}[1]{|{#1}\rangle}
\nc{\gb}{\bar{g}}
\nc{\sbar}{\bar{s}}
\nc{\Ab}{\bar{A}}
\nc{\Db}{\bar{D}}
\nc{\Lc}{\mathcal{L}}
\nc{\Oc}{\mathcal{O}}
\nc{\Qh}{\hat{Q}}

\begin{document}

\topmargin -5mm
\oddsidemargin 5mm

\setcounter{page}{1}

\vspace{8mm}
\begin{center}
{\huge {\bf A note on Kerr/CFT and free fields}}

\vspace{8mm}
 {\LARGE J{\o}rgen Rasmussen}
\\[.3cm]
 {\em Department of Mathematics and Statistics, University of Melbourne}\\
 {\em Parkville, Victoria 3010, Australia}
\\[.4cm]
 {\tt j.rasmussen@ms.unimelb.edu.au}

\end{center}

\vspace{8mm}
\centerline{{\bf{Abstract}}}
\vskip.4cm
\noindent
The near-horizon geometry of the extremal four-dimensional Kerr black hole
and certain generalizations thereof 
has an $SL(2,\mathbb{R})\times U(1)$ isometry group. Excitations around this 
geometry can be controlled by imposing appropriate boundary conditions.
For certain boundary conditions, the $U(1)$ isometry is enhanced to a 
Virasoro algebra.
Here, we propose a free-field construction of this Virasoro algebra. 
\renewcommand{\thefootnote}{\arabic{footnote}}
\setcounter{footnote}{0}

\newpage

\section{Introduction}

In the spirit of~\cite{BH86}, it was recently argued~\cite{GHSS0809} that 
the extremal four-dimensional Kerr black hole~\cite{BH9905} is holographically dual to a chiral 
conformal field theory (CFT) in two dimensions. 
The near-horizon geometry of the extremal four-dimensional Kerr black hole has 
an $SL(2,\mathbb{R})\times U(1)$ isometry group. 
Excitations around the near-horizon metric can be controlled by imposing
appropriate boundary conditions. To every consistent set of boundary conditions,
there is an associated asymptotic symmetry group generated by the diffeomorphisms
obeying the conditions. The conserved charge of an asymptotic symmetry is constructed as a
surface integral and can be analyzed using the formalism of~\cite{BB0111,BC0708}.
In the studies~\cite{GHSS0809,HMNS0811} of the Kerr/CFT correspondence and 
certain generalizations thereof based on the same isometry group, the $SL(2,\mathbb{R})$ 
isometries become trivial while the $U(1)$ isometry is enhanced to a Virasoro algebra. 

As generators of the asymptotic symmetry group, the Virasoro generators are
realized as differential operators. Our objective here is to construct a free-field
realization, resembling these differential operators, 
of the centrally-extended Virasoro algebra generated by the conserved charges. 
Our construction mimics the one introduced in~\cite{GKS9806} on string theory
on $AdS_3$ in relation to the $AdS$/CFT correspondence~\cite{Mal9711,GKP9802,Wit9802}. 
This construction was subsequently extended from the Virasoro 
algebra to various superconformal and topological conformal
algebras~\cite{Ito9811,YZ9812,And9901,YIS9902,Sug9909,AGS0002,Ras0002,Ras0003}, 
and we hope to discuss similar extensions of the present construction elsewhere.

Extending the work~\cite{Str9712} on the entropy of three-dimensional black holes such as the
BTZ black hole~\cite{BTZ9204}, it was found in~\cite{Car9812,Sol9812} 
that a copy of the Virasoro algebra appears in the near-horizon region of any black hole
and that it reproduces the black-hole entropy using the Cardy formula.
A free-field description arises in the approach of~\cite{Sol9812}, but it is not related
directly to the one proposed in the present work.

\section{Kerr/CFT correspondence}
\label{SecKerr}

In the global coordinates used in~\cite{GHSS0809}, the metric of the near-horizon
extremal Kerr black hole reads
\be
 d\sbar^2=2GJ\Omega^2\Big(\!-(1+r^2)d\tau^2+\frac{dr^2}{1+r^2}+d\theta^2
  +\Lambda^2\big(d\var+rd\tau\big)^2\Big)
\label{ds2Kerr}
\ee
where
\be
 \Omega^2=\Omega^2(\theta)=\frac{1+\cos^2\theta}{2},\qquad\qquad
  \La=\Lambda(\theta)=\frac{2\sin\theta}{1+\cos^2\theta}
\ee
Its isometry group is $SL(2,\mathbb{R})\times U(1)$, where
the rotational $U(1)$ isometry is generated by the Killing vector $\pa_\var$.

We are interested in perturbations $h_{\mu\nu}$
of the near-horizon geometry of the extremal black hole
whose background metric $\gb_{\mu\nu}$ is defined in (\ref{ds2Kerr}).
Asymptotic symmetries are generated by the diffeomorphisms whose action on the
metric generates metric fluctuations compatible with the chosen boundary conditions.
We are thus looking for contravariant vector fields $\eta$ along which the Lie derivative
of the metric is of the form 
\be
 \Lc_\eta \gb_{\mu\nu}\sim h_{\mu\nu}
\label{Lgh}
\ee
To such an asymptotic symmetry generator $\eta$, one associates~\cite{BB0111,BC0708}
the conserved charge
\be
 Q_\eta=\frac{1}{8\pi G}\int_{\pa\Sigma}\sqrt{-\gb}k_\eta[h;\gb]
   =\frac{1}{8\pi G}\int_{\pa\Sigma}\frac{\sqrt{-\gb}}{4}\eps_{\al\beta\mu\nu}d_\eta^{\mu\nu}[h;\gb]
     dx^\al\wedge dx^\beta
\label{Q}
\ee
where
\be
 d_\eta^{\mu\nu}[h;\gb]
  =\eta^\nu \Db^\mu h-\eta^\nu \Db_\sigma h^{\mu\sigma} +\eta_\sigma\Db^\nu h^{\mu\sigma}
  -h^{\nu\sigma}\Db_\sigma\eta^\mu+\frac{1}{2}h\Db^\nu\eta^\mu
  +\frac{1}{2}h^{\sigma\nu}\big(\Db^\mu\eta_\sigma+\Db_\sigma\eta^\mu\big)
\label{dmunu}
\ee
and where $\pa\Sigma$ is the boundary of a three-dimensional
spatial volume, ultimately near spatial infinity. Here, indices are lowered and raised
using the background metric $\gb_{\mu\nu}$ and its inverse, 
$\Db_\mu$ denotes a background covariant derivative, 
while $h$ is defined as $h=\gb^{\mu\nu}h_{\mu\nu}$.
To be a well-defined charge in the asymptotic limit, 
the underlying integral must be finite as $r\to\infty$.
If the charge vanishes, the asymptotic symmetry is rendered trivial.
The asymptotic symmetry group is generated by the diffeomorphisms 
whose charges are well-defined and non-vanishing.
The algebra generated by the set of well-defined charges is governed
by the Dirac brackets computed~\cite{BB0111,BC0708} as
\be
 \big\{Q_\eta,Q_{\hat\eta}\big\}
  =Q_{[\eta,\hat\eta]}+\frac{1}{8\pi G}\int_{\pa\Sigma}\sqrt{-\gb}k_{\eta}[\Lc_{\hat\eta}\gb;\gb]
\label{QQ}
\ee
where the integral yields the eventual central extension.

Written in the ordered basis $\{\tau,r,\var,\theta\}$, the boundary conditions considered
in~\cite{GHSS0809} are the fall-off conditions
\be
 h_{\mu\nu}=\Oc\!\left(\!\!\begin{array}{cccc} r^2&r^{-2}&1&r^{-1} \\ &r^{-3}&r^{-1}&r^{-2} \\ 
   &&1&r^{-1} \\ &&&r^{-1} \end{array} \!\!\right),
  \qquad\quad h_{\mu\nu}=h_{\nu\mu}
\label{hKerr}
\ee
and the zero-energy condition $Q_{\pa_\tau}=0$. Consistency of these conditions was
confirmed in~\cite{AHMR0906,DRS0906}.
The generators of the corresponding asymptotic symmetry group read
\be
 \xi=-\eps'(\var)r\pa_r+\eps(\var)\pa_{\var}
\label{xi}
\ee 
and form the centreless Virasoro algebra 
\be
 \big[\xi_{\eps},\xi_{\hat\eps}\big]=\xi_{\eps\hat\eps'-\eps'\hat\eps}
\label{Virdiff}
\ee
This symmetry is an enhancement of the exact $U(1)$ isometry generated by the Killing vector 
$\pa_\var$ of (\ref{ds2Kerr}) as the latter is recovered by setting $\eps(\var)=1$. 
The usual form of the Virasoro algebra 
is obtained by choosing an appropriate basis for the functions $\eps(\var)$ and $\hat\eps(\var)$.
With respect to the basis $\xi_n(\var)$, where 
\be
 \eps_n(\var)=-e^{-in\var}
\label{eps}
\ee
one introduces the dimensionless quantum versions 
\be
 L_n=\frac{1}{\hbar}\Big(Q_{\xi_n}+\frac{3J}{2}\delta_{n,0}\Big)
\label{L}
\ee
of the conserved charges. After the usual substitution $\{.,.\}\to-\frac{i}{\hbar}[.,.]$ of 
Dirac brackets by quantum commutators, the quantum charge algebra is 
recognized~\cite{GHSS0809} as the centrally-extended Virasoro algebra
\be
 \big[L_n,L_m\big]=(n-m)L_{n+m}+\frac{c}{12}n(n^2-1)\delta_{n+m,0},\qquad\quad
   c=\frac{12J}{\hbar}
\label{VirKerr}
\ee
This quantum charge algebra also arises when considering boundary conditions
sufficiently similar to those in (\ref{hKerr}).
A partial classification of such alternatives can be found in~\cite{Ras0908}. 

In~\cite{HMNS0811}, it was shown that the Kerr/CFT correspondence generalizes to a
Kerr-Newman-$AdS$-$dS$/CFT correspondence. The
near-horizon metric of these extremal black holes has an $SL(2,\mathbb{R})\times U(1)$
isometry group. Imposing the same boundary conditions as above, one obtains the same
asymptotic symmetry generators (\ref{xi}) enhancing the $U(1)$ isometry,
but a central extension depending on the parameters
characterizing the various generalizations of the extremal Kerr black hole.

\section{Free-field construction}

\subsection{Free fields}

Let us consider a pair of bosonic scalar fields $\var_1(z)$ and $\var_2(z)$ whose operator-product
expansions are given by
\be
 \var_i(z)\var_j(w)=-\delta_{ij}\ln(z-w)
\ee
The corresponding energy-momentum tensor and central charge are 
\be
 T_\var(z)=\sum_{i=1,2}\Big(-\frac{1}{2}\pa\var_i(z)\pa\var_i(z)-\frac{Q_i}{2}\pa^2\var_i(z)\Big),
   \qquad\quad  c_\var=2+3Q_1^2+3Q_2^2
\label{Tvar}
\ee
where $Q_i$ is the (renormalized) background charge associated to $\var_i(z)$.
Here and in the following, normal ordering is implied and regular terms are omitted.
We also consider a pair of bosonic ghost fields $\beta(z)$ and $\g(z)$ of conformal weight 
$\D(\beta)=\la$ and $\D(\g)=1-\la$, respectively, whose mutual 
operator-product expansion and corresponding energy-momentum tensor are given by
\be
 \beta(z)\g(w)=\frac{1}{z-w},\qquad
 T_{\beta\g}(z)=\la\beta(z)\pa\g(z)-(1-\la)\g(z)\pa\beta(z),\qquad
 c_{\beta\g}=12\la^2-12\la+2
\label{Tbetag}
\ee
We are mainly interested in the case where $\la=1$.

\subsection{Realization of Virasoro generators}

The objective here is to construct a free-field realization, resembling the differential-operator
realization $\xi_n$ in (\ref{xi}) with $\eps_n$ defined in (\ref{eps}), 
of the Virasoro algebra (\ref{VirKerr}). Our goal is thus to formulate a 
`direct quantization' of the differential operators reproducing the Virasoro algebra (\ref{VirKerr}). 
Mimicking the construction in~\cite{GKS9806}, 
we wish to realize the Virasoro generators as contour integrals like
\be
 L_n=\oint_0\frac{dz}{2\pi i}\Lc_n(z)
\label{Ln}
\ee
where $\Lc_n(z)$ is expressed in terms of free fields.
Invariance of this construction with respect to the free-field (`world-sheet') Virasoro generator 
\be
 T(z)=T_\var(z)+T_{\beta\g}(z),\qquad\quad T_n=\oint_0\frac{dz}{2\pi i}T(z)z^{n+1}
\ee
(here anticipating a construction in terms of $\var_1(z)$, $\var_2(z)$, $\beta(z)$ and $\g(z)$)
means that the two copies of the Virasoro algebra must commute
\be
 \big[T_n,L_m\big]=0
\ee
This is ensured if the integrands $\Lc_n(w)$ are
primary fields of conformal dimension 1
\be
 T(z)\Lc_n(w)=\frac{\Lc_n(w)}{(z-w)^2}+\frac{\pa\Lc_n(w)}{z-w}=\pa_w\Big(\frac{\Lc_n(w)}{z-w}\Big)
\label{TL}
\ee
which we will require in the following.

Initially following the standard procedure in Wakimoto free-field realizations~\cite{Wak86},
see~\cite{FF88,BF90,FF90,deBF9611,PRY9704,Ras9706} and references therein, 
we perform the `quantization'
\be
 r\to\g(z),\qquad\quad\pa_r\to\beta(z)
\label{rgrb}
\ee
where $\beta(z)$ and $\g(z)$ define a bosonic ghost system (\ref{Tbetag}) with $\la=1$.
The `residue' of the operator-product expansion $\beta(z)\g(w)$ thus reproduces
the commutator $[\pa_r,r]=1$. 

We likewise wish to replace the periodic coordinate $\var$ appearing in 
$\eps_n(\var)=-e^{-in\var}$ with a compactified bosonic scalar field $\var_1(z)$. However, 
exponentials like $e^{-in\var_1(z)}$ have $n$-dependent conformal properties
with respect to $T(z)$, in violation of (\ref{TL}). Furthermore,
the operator-product expansion of two exponentials like $e^{-in\var_1(z)}$ and 
$e^{-im\var_1(w)}$ can produce a pole or a zero of arbitrary order, preventing modes of the form
(\ref{Ln}) from generating a Virasoro algebra.
We thus propose to {\em screen} $\var_1(z)$ by introducing a companion bosonic
scalar field $\var_2(z)$ and consider
\be
 \var(z)=\var_1(z)+i\sigma\var_2(z),\qquad\quad \sigma^2=1
\label{var}
\ee
The main virtue, in this context, of the complex bosonic scalar field $\var(z)$
is its trivial operator-product expansion
\be
 \var(z)\var(w)=0
\label{varvar}
\ee
since it avoids the aforementioned obstacles faced by the use of $\var_1(z)$ alone.
{}From the present perspective of quantizing the differential operators, there is no apparent
reason for favouring $\var_2(z)$ compact instead of
non-compact. In order to interpret $\var(z)$ as a cylinder coordinate, though, one should
choose $\var_2(z)$ non-compact.
For convenience, we set $\sigma=1$.

Similarly to (\ref{rgrb}), the replacement of $\pa_\var$ should have `residue' 1 in its 
operator-product expansion with $\var(z)$. There is considerable freedom when choosing 
this replacement, due to (\ref{varvar}), but in anticipation of the invariance (\ref{TL}) imposed
below, we require that $\pa_\var$ is replaced by a linear
combination of $\pa\var_1(z)$ and $\pa\var_2(z)$. We thus supplement (\ref{rgrb}) by
\be
 \var\to\var(z),\qquad\quad \pa_\var\to-\pa\var_1(z)\ \ \big(\mathrm{mod}\ \pa\var(z)\big)
\ee

Combining the above, and with reference to (\ref{xi}) and (\ref{eps}),
our {\em ansatz} for the integrand $\Lc_n(z)$ now reads
\be
 \Lc_n(z)=ane^{-in\var(z)}\g(z)\beta(z)+e^{-in\var(z)}\Big(\!-b\pa\var_1(z)+b_n\pa\var(z)\Big)
\label{Lnabb}
\ee
where $a$, $b$ and $b_n$ are parameters to be fixed in the following.
Since
\be
 \oint_0\frac{dz}{2\pi i}e^{k\var(z)}\pa\var(z)=\delta_{k,0}\oint_0\frac{dz}{2\pi i}\pa\var(z)
\label{delta}
\ee
it follows that
\be
 L_n=\oint_0\frac{dz}{2\pi i}e^{-in\var(z)}\Big(an\g(z)\beta(z)-b\pa\var_1(z)\Big)
  +ib_0 p\delta_{n,0}
\label{b0}
\ee
where 
\be
 p=-i\oint_0\frac{dz}{2\pi i}\pa\var(z)=\oint_0\frac{dz}{2\pi i}\Big(\!-i\pa\var_1(z)+\pa\var_2(z)\Big)
\label{p}
\ee
This charge appears as (minus the usual convention for) the momentum generator of $\var(z)$ 
\be
 \var(z)=q-i(-p)\ln z+i\sum_{n\neq0}\frac{\al_n}{n}z^{-n}
\ee 
and it is noted that $p$ is a central element with respect to the Virasoro generators 
\be
 \big[L_n,p\big]=0
\ee

To verify that the ansatz (\ref{Lnabb}) does the job, and in the process fix the various parameters, 
one inserts the ansatz for $\Lc_n(z)$ into (\ref{VirKerr}) and (\ref{TL}). 
In the affirmative, one finds that the conserved charges $L_n$ in (\ref{Ln}) are invariant
(\ref{TL}) and generate the Virasoro algebra (\ref{VirKerr}) provided
\be
 a=1,\qquad\quad b=-i,\qquad\quad b_0=-\frac{3i}{2},\qquad\quad Q_1=Q_2=0
\ee
That is,
\be
 L_n=\oint_0\frac{dz}{2\pi i}e^{-in(\var_1(z)+i\var_2(z))}\Big(n\g(z)\beta(z)+i\pa\var_1(z)\Big)
  +\frac{3p}{2}\delta_{n,0}
\label{Lnp}
\ee
and the central charge is found to be
\be
 c=24p
\label{c24p}
\ee
It also follows that $p$ itself (and hence $c$) is invariant with respect to $T(z)$, and that 
the central charge of the free-field Virasoro algebra generated by $T(z)$ equals the number of
free fields, namely
\be
 c_\var+c_{\beta\g}=4
\label{cc4}
\ee

Now, comparing $c=24p$ in (\ref{c24p}) with the central charge $c=12J/\hbar$ associated to 
the near-horizon extremal Kerr black hole (\ref{VirKerr}), 
we see that the eigenvalues of the central element $p$ should be quantized according to
\be
 p\sim \frac{J}{2\hbar}
\label{pJ}
\ee
For given $J$, this imposes conditions on the representation theory of the free-field description,
much akin to the so-called Bose sea level~\cite{FMS86,VV88} of a bosonic ghost system.
Similar conditions, though based on different expressions for the central charge,
apply when considering the various generalizations of the extremal Kerr black hole
discussed in~\cite{HMNS0811} and referred to at the end of Section~\ref{SecKerr}.

Recalling the general bosonization recipe of~\cite{FMS86}, 
one may wonder if our construction in terms of the pair $\var_1(z)$ and $\var_2(z)$
corresponds to the bosonization of a bosonic ghost system 
\be
 \hat{\g}(z)\simeq e^{-i\var_1(z)+\var_2(z)},\qquad\quad
 \hat{\beta}(z)\simeq -e^{i\var_1(z)-\var_2(z)}i\pa\var_1(z),\qquad\quad \hat\la=1
\label{bghat}
\ee
independent from the one introduced in (\ref{rgrb}).
Since the associated $U(1)$ current $\hat\g(z)\hat\beta(z)$ is anomalous with
respect to $T_{\hat\beta\hat\g}(z)=\hat\beta(z)\pa\hat\g(z)$, while its 
bosonization $-\pa\var_2(z)$ is not (since $Q_2=0$), (\ref{bghat}) is not a `quantum equivalence'. 
This may not be of primary concern, though, since the only
given entities are the central charge and the differential operators in the classical limit.
The incompleteness of the bosonization (\ref{bghat}) nevertheless precludes invariance 
with respect to the corresponding free-field Virasoro generator
\be
 \hat T(z)=\beta(z)\pa\g(z)+\hat\beta(z)\pa\hat\g(z),\qquad\quad \hat c=4
\ee
As a remedy, one can modify the ansatz (\ref{b0}) and consider
\be
 L_n=\oint_0\frac{dz}{2\pi i}\hat{\g}^n(z)\Big(\hat a_n\g(z)\beta(z)
   +\hat b_n\hat{\g}(z)\hat{\beta}(z)\Big)+\hat b_n'p\delta_{n,0},\qquad\quad
    p=\oint_0\frac{dz}{2\pi i}\frac{\pa\hat{\g}(z)}{\hat{\g}(z)}
\ee
This is invariant and generates the Virasoro algebra provided $\hat a_n=n+1$, $\hat b_n=-1$
and $\hat b_n'=0$, that is,
\be
 L_n=\oint_0\frac{dz}{2\pi i}\hat{\g}^n(z)\Big((n+1)\g(z)\beta(z)
   -\hat{\g}(z)\hat{\beta}(z)\Big),\qquad\quad c=24p
\label{Lnp2}
\ee
where $[L_n,p]=0$.
In this picture, $p$ is related to the winding number of $\hat\g(z)$.
It is noted that, for $\hat\la=\frac{1}{2}$, the $U(1)$ current $\hat\g(z)\hat\beta(z)$ is anomaly free,
but $\hat\g^n(z)$ then has conformal dimension $n(1-\hat\la)\neq0$ with respect to the
corresponding free-field Virasoro generator.

It is emphasized that our proposal (\ref{Lnp}) is only one among many possible
quantizations of the differential operators underlying the Virasoro algebra appearing
in the (generalized) Kerr/CFT correspondence~\cite{GHSS0809,HMNS0811}. 
The somewhat related construction (\ref{Lnp2})
illustrates this assertion. Other possibilities arise when varying the parameter
$\la$ of the bosonic ghost system introduced in (\ref{rgrb}) as this changes the anomaly of 
the associated $U(1)$ current $\g(z)\beta(z)$ and in effect the conditions following from (\ref{TL}). 
One could also consider bosonizing this current (writing $\g(z)\beta(z)\simeq-\pa\var_3(z)$)
or to generalize the ansatz (\ref{Lnabb})
by allowing the coefficient $an$ to be general linear ($an\to an+n_0$), for example.
{}From a purely algebraic point of view, though, it appears that the proposal (\ref{Lnp}) is the one 
resembling the differential operators the most. Even so, there is no claim that this construction
actually describes the CFT appearing in the Kerr/CFT correspondence.
Better insight is gained once the corresponding representation theory is understood,
and we hope to address this theory and discuss correlation functions elsewhere.
Here, we merely point out that there is residual freedom when determining the vacuum
of the CFT for given central charge (eigenvalue of $24p$). It remains to 
be seen if a sensible pairing exists between the corresponding CFTs and the various 
generalizations of the extremal Kerr black hole considered in~\cite{GHSS0809,HMNS0811}.

It is desirable to test the (generalized) Kerr/CFT correspondence beyond the 
comparison~\cite{GHSS0809,HMNS0811} of the
black-hole entropy with the entropy computed using the Cardy formula. 
One such test was provided in~\cite{BHSS0907} on black-hole superradiance.
An alternative approach is to consider black-hole solutions of gravity theories
with higher-derivative corrections as discussed
in~\cite{SS9909,HHKT0805,KK0903,ACOTT0903} and references therein.
These higher-derivative terms would then affect the corresponding CFT,
in particular its central charge. 
One can hope that our construction provides means of addressing this issue.

\subsection*{Acknowledgments}
\vskip.1cm
\noindent
This work is supported by the Australian Research Council. 
The author thanks Omar Foda for discussions and encouragement,
and Oscar Dias, Chethan Krishnan and Sergey Solodukhin for comments.


\end{document}